\documentstyle[a4,12pt]{article}
\begin{document}
\begin{titlepage} \vspace{0.2in} \begin{flushright}
MITH-98/6 \\ \end{flushright} \vspace*{1.5cm}
\begin{center} {\LARGE \bf  Sonoluminescence Unveiled ?\\} \vspace*{0.8cm}
{\bf M.~Buzzacchi~,~~E.~Del Giudice~~and~~G.~Preparata}\\ \vspace*{1cm}
Dipartimento di Fisica dell'Universit\`a di Milano\\
and I.N.F.N. - Section of Milan\\
Via Celoria 16, 20133 Milan, Italy\\ \vspace*{1.8cm}

{\bf   Abstract:  \\ } \end{center} \indent
\baselineskip=18pt

The widening phenomenology of Single Bubble Sonoluminescence (SBSL), that is 
challenging generally accepted theory, is shown to be in good agreement 
with a new approach to condensed matter, based on the QED coherent 
interactions. The remarkable properties of SBSL are shown to emerge from the 
electromagnetic release of part of the latent heat of the water's 
vapour-liquid phase transition occurring at the bubble surface 
after it becomes supersonic. 
\baselineskip=12pt 
\vfill \begin{flushleft}  20 March 1998 \\
\end{flushleft}
\end{titlepage}
\baselineskip=24pt

Sonoluminescence (SL) or, better, its recent, thoroughly studied version
Single Bubble Sonoluminescence (SBSL) occupies a unique position in
contemporary physics \footnote{For a recent, very informative review see
\cite{report}}. On one hand its study does not require neither the
preparation of a complex, ill-characterized physical system nor the
construction of a very sophisticated and very expensive apparatus: this
has allowed many groups, within modest budgets, to carry out reproducible
experiments, thus establishing the undisputable physical reality of the
strange phenomenon whereby an acoustic field, through its pressure waves
in water and other liquids, is able to cause a gas bubble to emit light
up to maximum energy (about 6 eV) that can go, unattenuated, through
water. Furthermore, the remarkably short (less than 50 ps) and synchronous
(to within one part in 10$^{11}$) flashes emitted by the bubble during the
final stages of its collapse, add in a dramatic way, to the oddity of
the phenomenon. On the other hand, the consistent confirmations and
extensions of the original observations \cite{2} have posed a tremendous
challenge to the theoretical physics of our days, that so far has found
itself uncapable to give an explanation of the many oddities that SBSL
has been constantly exhibiting over the last seven years. In spite of
the many valiant theoretical efforts that have appeared in the literature,
the generally accepted conclusion today appears that the phenomena of
SL cannot be explained by known physics. Does this mean that the general
physical laws of the Standard Model need a correction or an extension?
We, like most of our colleagues, strongly believe that this is not, nor
can be the case. But, then, how can we understand the incredible properties
of the SL light flashes, both as to their photon energies and their time
coherence? The photon energies correspond to a blackbody temperature in
excess of 10$^5$ K, while the time coherence is in disagreement with both
the blackbody \cite{3},\cite{4},\cite{5} and the electron plasma mechanism 
\cite{5}.

In this Letter we present a theory of SL based on the general ideas of
QED, and its coherent interactions in condensed matter. The first
applications of this theory to a number of condensed matter systems have
been expounded in a recent book \cite{6}, and are also available in the
literature. Here, due to space limitations, we only mention that
the application to water \cite{7} 
\footnote{ As the theory has only been worked out for water, in this Letter 
we shall only deal with SBSL in water, but we find, in principle, no reason 
why, {\it mutatis mutandis}, a strictly parallel chain of arguments may 
not be developed for any system whose phase at room temperature is the liquid.}  
has shown that when a set of molecules
of H$_2$O reach a density about one third the density of ordinary water,
the system becomes unstable, and in a very short time (about 10$^{-14}$
s) it condenses into liquid water.
During the process of condensation, as we shall show and calculate below,
the excess energy -about 0.26 eV per molecule- gets electromagnetically
radiated in the surroundings with a typical, high energy spectrum.

Thus, the dynamical scenario that QED paints of SBSL is rather simple: when
in the final phase of collapse the surface of the "imploding" bubble reaches
a supersonic velocity with respect to the speed of sound in the water
vapour, there begins a process of compression which, when the vapour's
density reaches the value of 0.31 gcm$^{-3}$, leads to the formation of
liquid water in a very short time with the release of excess energy
(part of the latent heat) as a flash of light with a well defined 
energy spectrum,
both in frequency and intensity. The rest of this Letter deals with
the most relevant aspects of our theory.

According to the theory of QED coherence in matter~[6],[7], when the density 
of the water molecules $\rho$ is large enough ($\rho>\rho_{c}\simeq\frac{1}{3}
\rho_{Water}=10^{22}$~cm$^{-3}$) there exists a well defined excited state of 
the water molecule at the energy $\omega_{0}$=12.06~eV~\footnote{ Throughout this paper we 
employ natural units, where $\hbar=c=k_{B}=1$} that becomes the partner 
of the ground state in a coherent two-level oscillation in resonance with a 
coherent electromagnetic field whose spatial variation is determined by the 
wavelength $\lambda=\frac{2\pi}{\omega_{0}}$. This fact allows us to look 
at the system as an array of Coherence Domains (CD's), whose radii are $R_{CD}=
\frac{\pi}{\omega_{0}}\simeq500$~$\AA$, inside which the average (complex) 
amplitude of the electromagnetic field $A(\tau)$~~($\tau=\omega_{0}t$) obeys 
the short times equation~[6]:
\begin{equation}
\frac{i}{2}\ddot{A}~+~\ddot{A}~+~i\mu\dot{A} + g^{2}A~=~0,
\end{equation}
which when $\mu<$-0.5 ($g^{2}\ll$1, as it happens for H$_2$O~[7]) develops 
a run-away solution, i.e. an exponentially growing amplitude, signalling 
the transition toward a new state of the ensemble of molecules, the Coherent 
Ground State (CGS). Since the coefficient $\mu$ is proportional to 
$(N/V)^{1/2}$, it turns out that it is just at the critical density 
$\rho_{c}\simeq\rho_{water}$~[7], that $\mu$=-0.5, and the phase transition 
from the vapour to the liquid occurs. During the transition, the ensemble 
of molecules gains the energy $\delta E\simeq$0.26~eV per molecule and the 
time-scale for the transition is very short, of the order of 
$\frac{10^{2}}{\omega_{0}}\simeq10^{-14}$~sec. In a more complete and 
detailed work \cite{8} we show that the stationary state for both the matter 
and the electromagnetic field is reached after a number of "Rabi-oscillations" 
at the frequency $\omega_{R}\simeq1.1\omega_{0}$, with a characteristic 
damping time $t_{damping}=\frac{1}{\Gamma}$, where $\Gamma\simeq 1.5~\omega_{0}$.

In order to have an idea of how the water vapour inside the bubble may reach 
the critical density $\rho_{c}$, and give rise to the "implosive" condensation 
into liquid water predicted by QED coherence, we must look at the dynamical 
evolution of the gas bubble during the compression half-cycle of the sound 
wave. Without entering into a detailed analysis of the rather complicated 
hydrodynamics of the bubble, it is sufficient to note that at the time $t_0$, 
when the velocity of the bubble radius $R(t)$ [$R(t_{0})=R_{0}$] becomes 
supersonic with respect to the sound velocity in the gas, the system of 
gas molecules begins to be driven off thermodynamical equilibrium, the 
density inhomogeneities being unable to be leveled off through the propagation 
of (damped) sound waves. With respect to the "water-front" of the liquid 
proceeding toward the center of the bubble with increasing velocity, the 
gas molecules can be pictured as being (in the average) "frozen", and swept 
{\it in} 
by the imploding gas-liquid interface. At the time $t>t_{0}$ (See Fig.1) one 
expects that close to such interface there form a number of layers spaced by 
$a\simeq$3.2~$\AA$, the average distance between H$_2$O molecules in the 
liquid, while the transverse spacing $a_{T}(t)$ is given by:
\begin{equation}
a_{T}(t)=a_{0}\frac{R(t)}{R_{0}},
\end{equation}
where $a_{0}=a_{T}(t_{0})$ is the inverse of the third root of the vapour 
density $\rho_{0}$ at the time when the bubble becomes supersonic. In most 
treatments of the bubble collapse, at the time $t=t_{0}$ the gas temperature 
$T$ is close to the water boiling temperature (383~K at $p$=1.4~Atm), where 
the vapour density $\rho_{0}=\left(\frac{1}{a_{0}}\right)^{3}\simeq3\cdot
10^{19}$~cm$^{-3}$, and the number of H$_2$O molecules in the bubble volume 
is given by
\begin{equation}
N_{H_{2}O}=\frac{4\pi}{3}R_{0}^{3}\left(\frac{1}{a_{0}}\right)^{3}=1.14
\cdot10^{10}~~~(R_{0}\simeq4.5\mu m).
\end{equation} 

According to our theory, the bubble will continue to collapse until, at time 
$t=t^{*}$, the vapour density $\rho(t^{*})=\rho^{*}$ becomes
\begin{equation}
\rho^{*}=\frac{1}{a}\frac{1}{a_{T}(t^{*})^{2}}\simeq\frac{1}{3}
\left(\frac{1}{a}\right)^{3},
\end{equation}
and the transition from vapour to liquid occurs, thus liberating the SL 
flash. Note that at this time
\begin{equation}
a_{T}(t^{*})=a_{T}^{*}\simeq\sqrt{3}a=a_{0}\frac{R^{*}}{R_{0}},
\end{equation}
which setting $a_{0}\simeq3.2\cdot10^{-7}$~cm implies ($R_{0}\simeq4.5\mu m$)
\begin{equation}
R^{*}\simeq0.8\mu m.
\end{equation}

We can estimate the thickness of the condensing shell from the equation:
\begin{equation}
4\pi R^{*2}T\frac{1}{3}\left(\frac{1}{a}\right)^{3}\simeq
\frac{4\pi}{3}(R_{0}^{3}-R^{*3})\left(\frac{1}{a_{0}}\right)^{3},
\end{equation}
which stipulates the equality of the number of H$_2$O molecules contained 
in the spherical shell between $R_{0}$ and $R^*$ at the initial time 
$t=t_{0}$, with the number of molecules in the thin shell of radius $R^*$ 
when the critical density is reached. For $R_0=4.5\mu m$, $R^{*}\simeq0.8\mu m$ 
[See eq.(7)] and
\begin{equation}
T=\frac{R_{0}^{3}}{R^{*2}}\left(\frac{a}{a_{0}}\right)^{3}
\left[1-\frac{R^{*3}}{R_{0}^{3}}\right]\simeq1.4\cdot10^{-5}~cm,
\end{equation}
a bit over the size $\lambda$ of a coherence domain, and about 5.7 times 
smaller than the bubble radius $R^{*}$. A rather sensible result.

A detailed treatment of the electromagnetic release of the "latent heat" of 
the vapour-liquid transition~[8], shows that the classical electromagnetic 
current, associated with the oscillating two-level molecular systems, 
radiates the following e.m. spectrum per unit volume:
\begin{equation}
\frac{1}{V}\frac{dE}{d\omega}=
\frac{3\omega_{0}^3}{16\pi^{3}}|c_{1}|^{2}
|F(\omega)|^{2}\frac{\omega^2}{(\omega-\omega_{R})^{2}+
\frac{\Gamma^{2}}{4}},
\end{equation}
where $|F(\omega)|=\exp\left(-1.4\left(\frac{\omega}{\omega_{0}}\right)^{2}
\right)$ is the form factor of the water CD's, and the width $\Gamma\simeq
18$~eV is determined through the requirement that the total energy output 
equals 0.26~eV per molecule. The constant $|c_{1}|^{2}$ has the value 1.8. 
The complete spectrum [Eq.(9)] is reported in 
Fig.2, and a comparison with a typical experimental spectrum is shown in 
Fig.3. 

Due to the water opacity for $\omega>$6~eV, only a small fraction of 
the 0.26~eV per molecule, 
about 0.03~eV, gets detected; the rest, which extends up to the large energy 
values 15$\div$20~eV, is either absorbed in water or by the gas molecules, thus 
rendering the bubble a unique, high energy chemical reaction chamber, as 
recently noted in Ref.\cite{9}. This observation appears particularly 
significant, since it clearly shows the important role of noble gases, 
whose excitation (ionization) energies lie above (or close to) the cut-off 
energy of the emitted spectrum. This means that these gases will be largely 
unaffected by the SL e.m. flash and their dynamical evolution is sensitive 
only to the hydrodynamics of the process, hence very stable from cycle to 
cycle. The situation is completely different for diatomic molecules, such as 
the $N_{2}$'s or $O_{2}$'s af air; their transformation into highly interactive 
free radicals will lead to their disappearance from the bubble, as recently 
noted~[9]. This fact may also explain why for purely diatomic gases SL is 
so unstable and shows very strange memory effects~[1].

As for the recent, very interesting observations of the temporal widths of the 
SL flashes, they seem to be in fair agreement with our theory, which predicts 
that:
\begin{enumerate}
\item all frequencies are emitted simultaneously, there being no delay effect 
for the low frequencies with respect to high frequencies;

\item the temporal widths do not depend on the nature of the bubble gases;

\item the actual values of the widths depend on the deviations of the 
imploding interface from sphericity, causing its different parts to reach 
the critical density at different times. Please note that giving such 
fluctuations the size of a CD (a rather reasonable assumption), i.e. 
$\lambda\simeq10^{-5}$~cm, and considering a radius velocity at $R=R^{*}$ 
about 1.5$\cdot10^{5}$ cm/s, yields $\Delta t\simeq$60~ps, of the right 
size.
\end{enumerate}

Finally, regarding the coherence of the SL flashes, our theory predicts 
good coherence properties, for the observed e.m. radiation originates from 
the classical currents associated with the two-level oscillations of the 
large number of H$_2$O molecules of the different ($\simeq$150)~CD's.

Leaving a more detailed analysis to a future publication~[8], we would 
like to stress that in this Letter we have tried to present the main 
conceptual paths that bring the general theory of QED to bear upon the 
fascinating phenomena of SL. The good, not only qualitative, picture that 
our theory yields of the widening SL phenomenology shows, in our opinion, 
that we now possess a rather adequate and sufficiently precise description 
of the dynamics of water and of its formation from a disordered assembly of 
vapour molecules, when their density reaches the threshold  
$\rho_{c}$. The association of this event with a remarkable burst of e.m. 
radiation of rather high frequency, whose spectrum and intensity we correctly 
predict, bears witness, we believe, to the basic electrodynamical nature of 
the gas-liquid phase transition of H$_2$O, which is at variance with the 
generally accepted views about water and condensed matter in general.

\newpage
\begin{center}
{\LARGE \bf Figure Captions}
\end{center}
\vskip1cm
{\bf Figure 1}: The relevant parameters of the collapsing bubble: $R_{0}=$ 
radius at which the compression becomes supersonic. $R^{*}=$ radius at which 
the sonoluminescence burst is emitted.
\vskip1cm
{\bf Figure 2}: The energy spectrum $dE/d\omega$ for one molecule; due to the 
water's opacity only the portion $\omega<$5~eV of the spectrum is detected.
\vskip1cm
{\bf Figure 3}: Comparison of the energy spectrum $dE/d\lambda$ irradiated in 
each sonoluminescence pulse with the experimental data.


\begin{thebibliography}{99}
\bibitem{report} B.P.~Barber, R.A.~Hiller, R.~L\"{o}fstedt, S.J.~Putterman, 
K.R.~Weninger, Phys. Reports {\bf 281}, 65-143 (1997).

\bibitem{2} B.P.~Barber, S.J.~Putterman, Nature {\bf 352}, 318 (1991).

\bibitem{3} H.~Metcalf, Science {\bf 279}, 1322 (1998).
 
\bibitem{4} R.A.~Hiller, S.J.~Putterman, K.R.~Weninger, Phys.~Rev.~Lett. 
{\bf 80}, 1090 (1998).

\bibitem{5} B.~Gompf, R.~G\"{u}nther, G.~Nick, R.~Pecha, W.~Eisenmenger, 
Phys. Rev. Lett. {\bf 79}, 1405 (1997).

\bibitem{6} G.~Preparata, {\it QED Coherence in Matter}, World Scientific, 
Singapore, 1995.

\bibitem{7} R.~Arani, I.~Bono, E.~Del Giudice, G.~Preparata, Int.~J.~Mod.~
Phys. {\bf B~9}, 1813 (1995).

\bibitem{8} M.~Buzzacchi, E.~Del Giudice, G.~Preparata (in preparation).

\bibitem{9} T.J.~Matula, L.A.~Crum, Phys. Rev. Lett. {\bf 80}, 865 (1998).
\end{thebibliography}
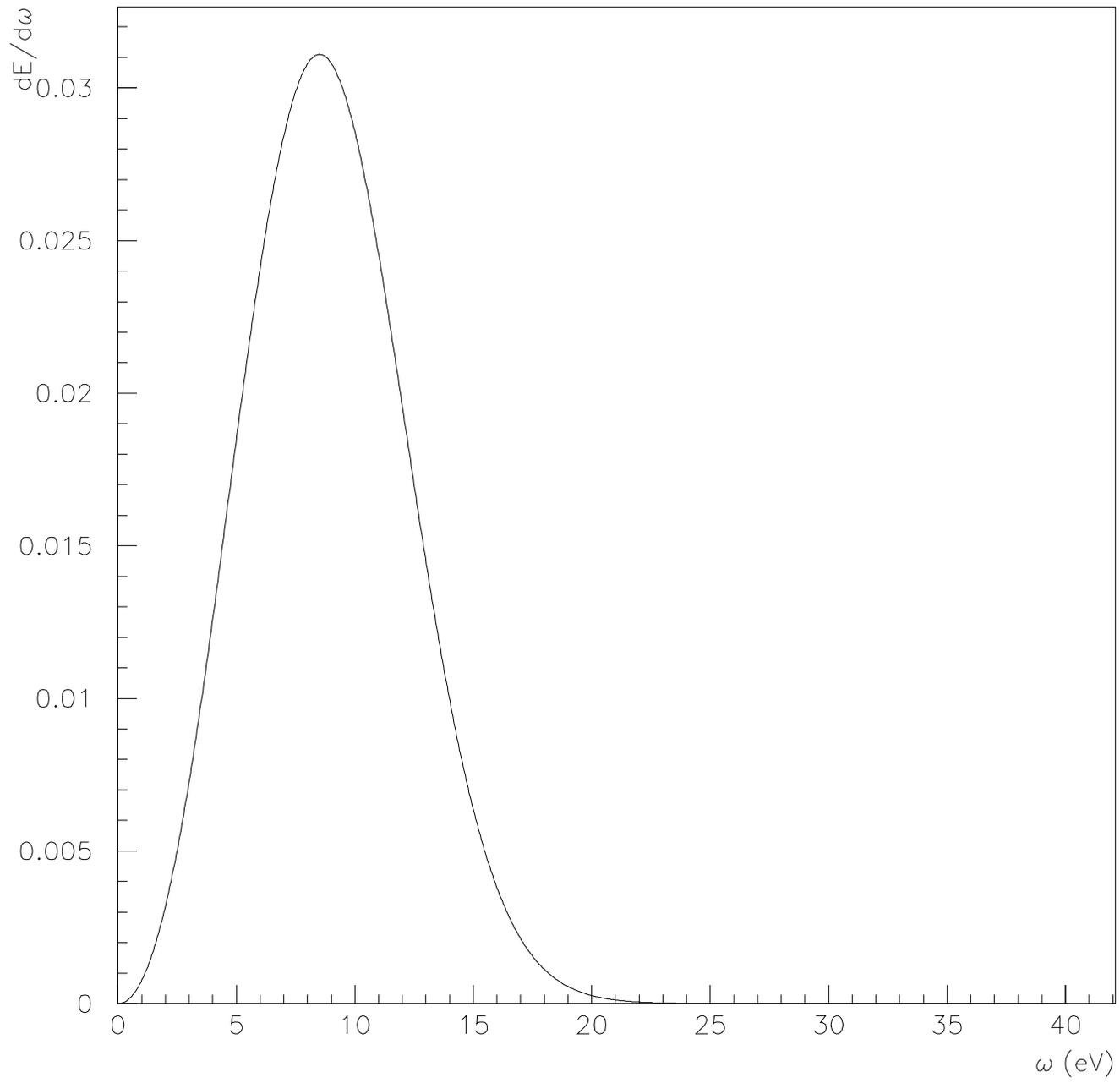
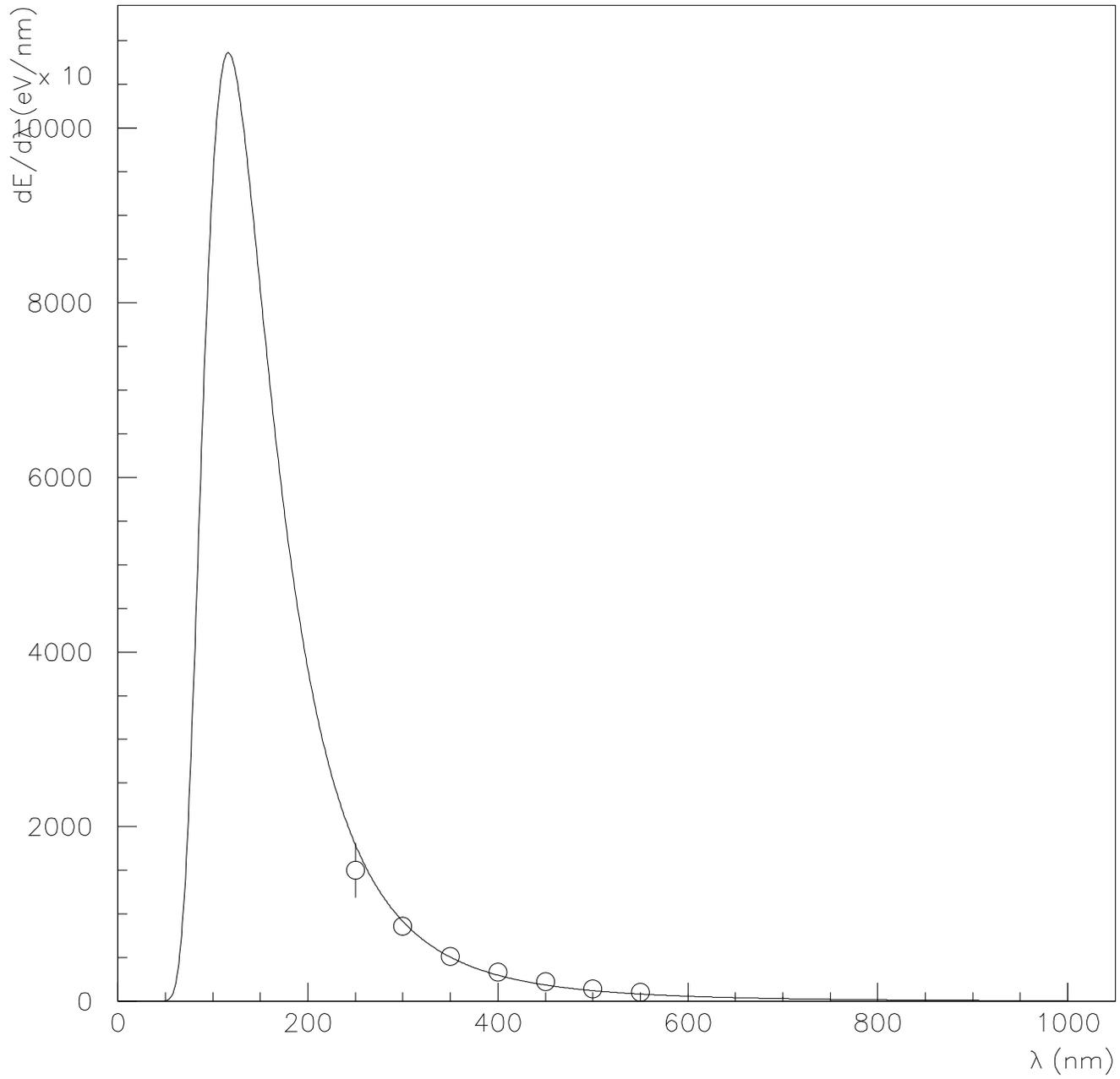
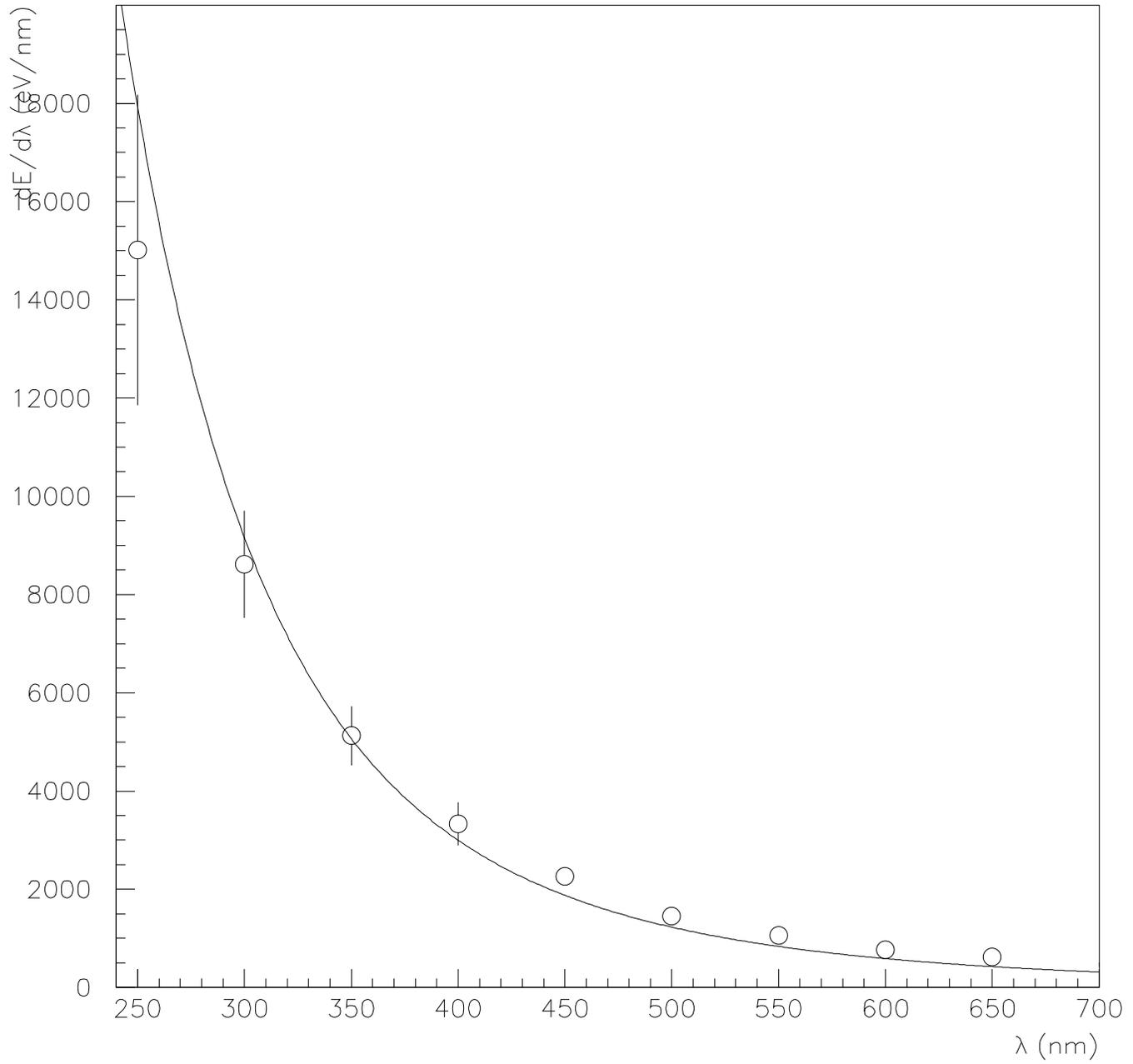
\end{document}